\documentclass[12pt]{article}
\usepackage{amssymb,2up}

\input cteq5.pre		

\begin{document}

\input front.inp        

\section{Introduction}

The structure of hadrons represented by parton distributions is an essential
part of our knowledge of the elementary particle physics world. The
interpretation of existing experimental data in terms of the Standard Model
(SM), the precision measurements of SM parameters, as well as the direct
search for signals for physics beyond the SM, all rely heavily on
calculations based on Quantum Chromodynamics (QCD) and the QCD-parton
picture, with the parton distribution (and fragmentation) functions as 
essential input. The (non-perturbative) parton distribution functions at
some given momentum scale are currently determined phenomenologically by a
global analysis of a wide range of available hard scattering processes
involving initial-state hadrons, using the perturbative QCD-parton framework.

The global analysis of parton distributions requires a continuing effort. As
new experimental and theoretical advances occur, the parton distributions
can be determined with increasing accuracy and assurance. Although current
knowledge of the parton distributions, based on several generations of
global analyses, is far more quantitative than in the early years of the
parton model, gaps still remain, as will be discussed at the end of this
paper. In addition to filling these gaps, there are two important
motivations for the vigorous pursuit of global analysis: (i) a comprehensive
global analysis constitutes an important test of the consistency of
perturbative QCD when available experimental constraints exceed the degrees of 
freedom inherent in the non-perturbative shape parameters, 
and provides a powerful tool to discover the boundaries of
applicability of the conventional theory, hence to discover hints for
the need of new physics tools or ideas; and (ii) since the global analyses
inevitably involve both experimental and theoretical uncertainties, it is
important to quantify the uncertainties in the resulting parton
distributions. Some important efforts on precision measurement of SM
parameters such as the W mass, as well as the determination of signals and
background for new physics searches, are limited by uncertainties on parton
distributions.

This paper extends the series of global QCD analyses of the CTEQ group 
\cite{CTEQ123,CTEQ4} to include significant new experimental results of the 
last two years. Sec.\ \ref
{sec:Expts} summarizes these new experimental developments. Sec.\ \ref
{sec:GlobalIssue} discusses issues that arise in a quantitative global QCD
analysis of these data, and our approach to these issues. Sec.\ \ref
{sec:CTEQ5} describes the main results of our analysis, in the form of
several sets of new CTEQ5 parton distributions, chosen to meet the needs of
different types of applications as required by the consistent use of QCD
theory. In Sec.\ \ref{sec:Compare} we compare our results with those of
recent parallel efforts and point out the origin of the differences.
Finally, in Sec.\ \ref{sec:Conclude} we describe the remaining open problems
in the global analysis of parton distributions, as well as various sources
of uncertainties in the parton distribution parameters and the prospect for
quantifying the uncertainties. In the Appendix we summarize some issues
regarding the choice of renormalization and factorization schemes for the
treatment of heavy quarks, which are relevant for quantitative global analysis
of increasingly precise data in many processes.


\section{New experimental information and their use in global analysis}

\label{sec:Expts}

Since the publication of the last round of CTEQ global QCD analysis \cite
{CTEQ4}, improved and new experimental data have become available for many
processes. These are summarized here. The use of these data in the global
analysis depends on theoretical considerations which we discuss in the next
section.

\textbf{Deep inelastic scattering}: The NMC and CCFR collaborations have
finished and published analyses of their respective data on 
muon-nucleon \cite{NMC}
and neutrino-nucleus \cite{CCFR} scattering. These new results lead to
subtle changes in their implications for $\alpha _{s}$ and parton
distribution determination. The H1 and ZEUS collaborations at HERA have
published more extensive and more precise data on the total inclusive
structure function $F_{2}^{p}$ \cite{H1f2,ZEUSf2}.%
\footnote{New measurements of the structure function $F_2$ from the 95-96 
HERA run have not yet been made available for global analysis.} 
These results provide
tighter constraints on the quark distributions, as well as on the gluon
distribution, mainly through the $Q$-evolution of the structure functions.
The HERA experiments also present new data on semi-inclusive $F_{2}^{c},$
with charm particles in the final state \cite{H1C,ZEUSC}. The analysis of
the $F_{2}^{c}$ data will be discussed in the next section.

\textbf{Lepton-pair production ($p/d$) asymmetry}: The E866
collaboration has measured the ratio of lepton-pair production (Drell-Yan
process) in $pp$ and $pd$ collisions over the $x$ range 0.03 -- 0.35 \cite
{E866}, thus expanding greatly the experimental constraint on the ratio of
parton distributions $\bar{d}/\bar{u}$ (compared to 
the single point of NA51 at $%
x=0.18$ \cite{NA51}). This data set has the most noticeable impact on the
new round of global analysis.

\textbf{Lepton charge asymmetry in W-production}: The CDF collaboration has
improved the accuracy and extended the $y$ range of the measurement of the
asymmetry between $W\rightarrow\ell^{\pm}\nu$ at the Tevatron \cite{CDFlasy}. 
This provides additional constraints on $d/u$.

\textbf{Inclusive large $p_{T}$ jet production}: The D0
collaboration has recently finished the final analysis of their 
inclusive jet production data, including information on the correlated
systematic errors \cite{D0Jet}. The CDF collaboration also has presented new
results from their RunIB data set \cite{CDFIB}. Systematic errors in these
data sets dominate the experimental uncertainty over much of the measured $%
p_{T}$ range. The correlated systematic errors provide important information
on the shape of the differential cross-section, $d\sigma /d p_T$,
and constrain the parton distributions accordingly.

\textbf{Direct photon production}: The E706 collaboration at Fermilab has
published the highest energy fixed-target direct photon production data
available to date \cite{E706}. The measured cross-sections lie a factor of $%
2-3$ above the traditional next-to-leading (NLO) QCD calculation, thus
posing a real challenge for their theoretical interpretation and their use
in global analysis. 


\section{Global analysis issues and procedures}

\label{sec:GlobalIssue}

In this section, we consider various physics issues relevant to
incorporating the new experimental data in the global analysis of parton
distribution functions, and describe the specific inputs to the CTEQ5
analysis.

\textbf{Charge asymmetry data and quark flavor differentiation}:

Most inclusive processes are not sensitive to differences between the quark
parton flavors, since contributions from them are summed in the
cross-section. In global analysis, these differences represent ``fine
structure'' that can be resolved by including physical quantities asymmetric
in the various flavors. In particular, the difference between the $u$ and $d$
quarks is determined by differences between cross-sections with
proton/neutron targets in DIS and Drell-Yan processes, or with $W^{\pm} $
final states (manifested by the decay leptons) in $\bar{p}p$ collisions. As
mentioned in the previous section, new data from E866 and CDF have an
immediate impact on flavor differentiation in current global analyses. \
These new data are complemented by the final results from the very precise
measurement of $F_{2}^{d}/F_{2}^{p}$ by the NMC experiment.

In our analysis, information from the E866 $\sigma _{pd}/\sigma _{pp}$
Drell-Yan experiment is maximized by treating the data sets above and below
the $\Upsilon $ peak separately (rather than integrating over the invariant
mass of the lepton pair, with an $x$-dependent $\Upsilon $ gap).%
\footnote{%
Data in the $\Upsilon $ region are excluded since they involve different
physics. We thank Paul Reimer of E866 for providing the detailed
information on the measurement in the two separate regions which makes this
treatment possible.} For the CDF W-lepton asymmetry data, we use the
resummed NLO calculation provided by the ResBos program \cite{Balazs:1997xd}.%
\footnote{%
We thank Csaba Balazs for assistance
in this calculation.} Resummation has an effect on the theoretical
calculation at large rapidity of the lepton, due to experimental cuts on the
lepton $p_{T}$.

An important source of uncertainty in the study of quark flavor dependence
arises from the necessity of using DIS and Drell-Yan data on a deuteron
target, in lieu of a neutron target. On general grounds, the impulse
approximation of considering the deuteron cross-section as the incoherent
sum of those of proton and neutron is expected to be good at small $x.$ In
the large $x$ region, there have been studies of the ``deuteron correction''
factor needed to extract the neutron cross-section from deuteron data; but
there is no universally accepted theory on the size and shape of this
correction. Furthermore, there are additional complications in the large $x$
region such as ``higher-twist'' effects of various origins, including target
mass effects. Recently, these problems have been revisited by two
phenomenological studies \cite{Slac,BodekYang}. Using approximate (i.e.\
modified MRSA) quark-distributions, Ref.\ \cite{BodekYang}\ advanced the
case for an unconventional behavior of the $d/u$ ratio at large $x,$ as the
result of specific deuteron- and target-mass corrections; and pointed out
the importance of studying this issue in a full global QCD analysis. In 
performing such an analysis, we have
found that equally consistent descriptions of all
current data can be obtained with or without applying the corrections of 
\cite{Slac,BodekYang} to the DIS deuteron cross-sections. Consequently, for
the general purpose CTEQ5 parton distributions, to be used in usual
applications, we follow the conventional practice of treating deuteron
cross-sections as an incoherent sum of proton and neutron ones. A more
specific study of this problem, including recent theoretical development of
higher twist effects and possible ways to distinguish between the
alternative behaviors of the $d/u$ ratio at large $x$ in HERA measurements,
will be presented in a forthcoming paper.

Neutral current (NC) and charged current (CC) DIS scattering (initiated 
by charged leptons and neutrinos respectively) are sensitive to different
combinations of quark flavors, hence also provide information on their
differences. Here one encounters a separate set of inter-related 
uncertainties: heavy-target correction for neutrino experiments, 
origin of the apparent disagreement of measured ratio of
CC and NC structure functions with the classic ``charge ratio'' 
($5/18$ rule) in the region $x<0.1$, the strange quark fraction, and the
validity of charge symmetry for parton distributions 
(i.e.\ $f_p^u = f_n^d$ \dots). For
some recent investigations, see \cite{ChargeSym}. These open questions 
deserve further study. In the absence of compelling reasons to do otherwise, 
we follow the practices of previous CTEQ (and MRS) analyses 
\cite{CTEQ4,CTEQ123} on these issues.

\textbf{Direct photons, inclusive jets, and the gluon distribution}:

Since the recently published E706 direct photon data \cite{E706}, measured
at 530 and 800 GeV, cover a wide range of $x,$ and report comparatively
small statistical and systematic errors, one might hope to determine the
gluon distribution directly from this process over the full range covered by
this and earlier experiments. However, the measured cross-section by E706 is
roughly a factor of $2-3$ larger than the conventional NLO QCD calculation. This
result strengthens a previous suggestion \cite{CTEQdph} that initial state
parton $k_{T}$ broadening, due to multiple soft-gluon radiation, greatly
enhances the steeply falling photon $p_{T}$ spectrum \cite{E706}. However,
to make use of these data to determine the gluon distribution with any
confidence, one needs a theory capable of predicting the needed
large theoretical correction factor with considerable accuracy. Does that theory
exist?  In a recent paper \cite{CteqE706}, it is shown that a
phenomenological treatment of the $k_{T}$ broadening effect based on
conventional Gaussian smearing, with the amount of smearing determined by
data from related processes such as di-photon and photon+jet, can
consistently describe a wide range of existing data on direct photon and
related pion production. However, this study also reveals the expected fact
that the results of such calculations are still rather model-dependent. A
variety of uncertainties associated with the choices of parameters,
phenomenological procedures, scales, ... etc. described in this study show
that the shape as well as normalization of the $p_{T}$ spectrum can be
significantly affected by choices made in the model calculation.%
\footnote{%
One revealing fact from Ref.\ \cite{CteqE706} is that our model results
differ from those of Ref.\ \cite{MRST}, which uses a different
phenomenological procedure, in both normalization and shape of the $p_{T}$
spectrum, for a similar nominal amount of $k_{T}$ broadening. Although this
difference does not measure any meaningful ``uncertainty'', the implication
about the inherent ambiguity in using direct photon data to determine the
gluon distribution is obvious.} The wide range of uncertainty is underlined
by another recent study \cite{Aurenche98}, which argues that aside from
E706, $k_{T}$ broadening is not necessarily required to reconcile NLO QCD
and earlier fixed-target and ISR direct photon experiments. 

In short, one finds that the consistency between existing fixed-target
experiments on
direct photon is still open to question, the problem being partly dependent
on the theoretical framework used to compare experiments at different
energies. Furthermore, the QCD theory for direct photon production in the
$p_T$ range of these experiments is very much in a state of
active development: \ Are resummation effects beyond NLO as large as a
factor of 2$\thicksim$3 at E706 energies? \ How quantitative can the
resummation theory become?  We refer the reader to Ref.\ \cite
{CteqE706,Aurenche98,DphResum} for detailed discussions on these issues.
Under the present circumstances, it is impossible to incorporate these 
experiments in the global QCD analysis without introducing subjective
choices of experimental data, as well as model-dependent theoretical
procedures.
We note that, direct photon production has also been measured at 
hadron colliders \cite{DirPhCollider}. The cross-section at high $p_T$ agree
rather well with NLO QCD calculations; however, at the low $p_T$ end, one also 
observes an enhenced cross-section compared to theory. The statistics for
these experiments are currently too low to make these data useful for the 
global analysis of parton distributions.

Inclusive large $p_{T}$ jet production at the Tevatron, on the other hand,
provides a much more reliable experimental constraint on the gluon
distribution, since the NLO QCD theory has been shown to be rather stable 
\cite{EKS,GGK} in the region $p_{T}>40$ GeV where measurements exist. 
This energy scale is considerably higher than that of fixed-target direct
photon discussed in the previous paragraph. Multi-soft gluon effects 
are insignificant for data in this range. Preliminary CDF and D0 
data were used in the previous CTEQ4 analysis. It was shown that for the
determination of the gluon distribution, the inclusive jet data supplement
very well the precise, but indirect, constraints implied by the $Q$%
-dependence of DIS data. Now that these experimental results 
have been finalized \cite{D0Jet,CDFIB}, 
it is natural to take full advantage of them in the new
global analysis. An exhaustive study, based on all available data, confirms
the previous finding that the combination of jet and DIS data constrain the
gluon distribution quite well in the range $0.05<x<0.25$. We will test the
parton distributions obtained in this way against the direct photon data
to see whether we get a consistent picture of the gluon.

\textbf{The strong coupling $\alpha_{s}$}:

For this study, we have kept the value of $\alpha_{s}(m_Z)$ fixed at $0.118$. 
As is well known, the value of $\alpha _{s}$ is strongly
coupled to the gluon distribution in analyzing lepton-hadron and
hadron-hadron processes. In particular, this correlation has been examined
in Ref.\ \cite{CTEQ4}, and presented in the ``A-series'' CTEQ4 parton
distributions which span a range of $\alpha _{s}$ around the world average.
\ When $\alpha _{s}$ is left as a free parameter in the current fit, we find
a range of values of $\alpha _{s}(m_{Z}),$ including $0.118,$ which give
almost equally good fits. Since
there are many more constraints on $\alpha _{s}$ from processes beyond those
used in parton distribution analysis, many of which are independent of the
uncertainties on the gluon distribution, we have chosen to use a fixed $%
\alpha _{s}(m_{Z})$ in the final CTEQ5 analysis. We will comment on the
effect of varying $\alpha _{s}$ in the next section. The range of variation
of parton distributions due to a variation in $\alpha _{s}$ can still be
inferred from Ref.\ \cite{CTEQ4}.

\textbf{Charm production in DIS:}

Preliminary results on charm production at HERA \cite{H1C,ZEUSC} have
highlighted the need for a more careful treatment of heavy quarks in the
perturbative QCD (PQCD) formalism. Although theories for heavy quark
production exist \cite{GRStrat,SBLeiden,ACOT,Collins98}, and the CTEQ4 analysis
provided several sets of parton distributions which incorporate charm quark
mass effects (CTEQ4HQ,4F3,4F4), it is important to bear in mind the
limitations of the current state of the art on this subject. Experimentally,
one can only measure the cross-sections for producing $D$ and $D^{\ast }$
mesons in certain kinematic ranges. Extracting $F_{2\text{,exp}}^{c}$
requires: (i) an extrapolation of $D$- and $D^{\ast }$-production data to
the full phase space to obtain $F_{2}^{D,D^{\ast }}$; and (ii) a procedure
to infer $F_{2\text{,exp}}^{c}$ from $F_{2}^{D,D^{\ast }}$ involving, among
other uncertainties, the not so well-known fragmentation functions for $%
D,D^{\ast }$. On the theoretical side one faces a different, but related,
dilemma. On one hand, among the existing schemes for treating heavy quarks
in PQCD, $F_{2\text{,th}}^{c}$ is in principle defined only in the
fixed-three-flavor scheme (with $u,d,s$ being the only quark partons); but
this scheme is not suitable for quantitative treatment of high energy (i.e.\
collider) inclusive processes which are essential for global QCD analyses.
On the other hand, in generalized \msbar{} schemes admitting a non-zero mass
charm quark as an active parton at high energies\footnote{%
Some background information on the available theoretical schemes for
treating heavy quarks in PQCD, useful for this discussion and that of the
following section, is provided in the Appendix.}, which are suitable for
global analyses at high energies, ``$F_{2}^{c}$'' is not a well-defined
quantity in principle because the naive ``$F_{2}^{c}$'' contains large
logarithms of the same type that are resummed into charm parton
distributions. Only in $F_{2}^{\text{tot}},$ and in $F_{2}^{D,D^{\ast }},$
do these logarithms cancel between contributions from all parton flavors to
yield infrared safe quantities that are suitable for comparison with
experiment. For these experimental and theoretical reasons, the emerging
charm production measurements can best be used as a testing ground for
further development of both, rather than as a mature input to quantitative
parton distribution analysis.\footnote{%
The situation is somewhat similar, but not identical, to that of jet physics
before the development of practical and infrared safe jet algorithms that
allow a meaningful comparison between theory and experiment.} \ Thus, we
shall not use the preliminary data on charm production in DIS in the global
analysis; but we will present a comparison of the new parton distributions
with available data, using order $\alpha _{s}$ formulas in the generalized
4-flavor \msbar{} scheme.

\section{CTEQ5 Parton Distributions}

\label{sec:CTEQ5}

Based on the considerations discussed above, we carried out an extensive
round of global analyses, using DIS data sets from BCDMS \cite{BCDMS}, NMC 
\cite{NMC}, H1 \cite{H1f2}, ZEUS \cite{ZEUSf2}, CCFR \cite{CCFR}, E665 \cite
{E665}; Drell-Yan data from E605 \cite{E605}, E866 \cite{E866};
W-lepton-asymmetry data from CDF \cite{CDFlasy}; and inclusive jet data from
D0 \cite{D0Jet} and CDF \cite{CDFIB}. The kinematic ranges spanned by the
various experiments, and the wide scope of the overall coverage, are shown
in Fig.\ \ref{fig:KinMap}. (Data points below $Q^2=4$ GeV$^2$ are not included
in this twist-two QCD analysis; hence are absent on this plot.)
The complementary roles of the fixed-target, HERA, and
Tevatron experiments are clearly illustrated in this plot.

Various theoretical and
phenomenological issues described in the previous section are explored by
systematically studying the effect of reasonable variations of the known
uncertainties in each case. 
The initial (non-perturbative) parton distributions are
parametrized at $Q_{0}=1\,$GeV; cuts on the kinematic variables $%
Q,W,p_{T}\dots$ on data points used are generally of the order of 2 -- 4
GeV, the same as in \cite{CTEQ4}. Our results are not sensitive to the
specific choices.

The following sets of CTEQ5 parton distributions, representative of our best
fits, are provided for general purpose use in applications of perturbative
QCD to calculate high energy processes, as well as for special purpose
applications as specified. These sets are summarized in Table I.%
\footnote{
Fortran computer codes for these parton distribution functions can be 
downloaded from the Web site http://cteq.org.
}

\begin{center}
\begin{tabular}{|cc|}
\hline
PDF set & \multicolumn{1}{|c|}{Description} \\ \hline\hline
& conventional (zero-mass parton) sets \\ \hline
CTEQ5M & \multicolumn{1}{|c|}{\msbar\ scheme} \\ \hline
CTEQ5D & \multicolumn{1}{|c|}{DIS scheme} \\ \hline
CTEQ5L & \multicolumn{1}{|c|}{Leading-order} \\ \hline\hline
CTEQ5HJ & \multicolumn{1}{|c|}{large-$x$ gluon enhanced} \\ \hline\hline
& on-mass-shell heavy quark sets \\ \hline
CTEQ5HQ & \multicolumn{1}{|c|}{\msbar\ (ACOT) scheme} \\ \hline
CTEQ5F3 & \multicolumn{1}{|c|}{fixed-flavor-number ($N_{f}=3$) scheme} \\ 
\hline
CTEQ5F4 & \multicolumn{1}{|c|}{fixed-flavor-number ($N_{f}=4$) scheme} \\ 
\hline
\end{tabular}
\end{center}

\textbf{CTEQ5 sets in the conventional schemes:}

The \textbf{CTEQ5M} set is defined in the \msbar\ scheme, matched with
conventional NLO hard cross-sections calculated in the zero-quark-mass
approximation for all active flavors, including charm and bottom. This set
is the most convenient one to use for general calculations, as the vast
majority of available hard cross-sections in the literature and in existing
programs have been calculated in this limit. It represents an updated
version of the CTEQ4M distribution set.

\textbf{CTEQ5D} is the corresponding set in the DIS scheme, obtained by an
independent fit (rather than making a theoretical transformation from
CTEQ5M),\footnote{%
Performing a transformation from one scheme to another by the order $%
\alpha_{s}$ perturbative formula \cite{MRST2} can lead to large errors for
the gluon and the sea quarks in kinematic regions of $(x,Q)$ where the
``leading order'' term is small compared to the correction term: e.g.\
valence quark corrections to the gluon, and gluon correction to the sea
quarks at large $x$.} using identical experimental input and fitting
procedure. Likewise, \textbf{CTEQ5L} is the corresponding set in leading
order QCD, which should be appropriate for simple calculations and for use
in many Monte Carlo programs.

Fig.\ \ref{fig:AllPdf} shows an overview of the parton distributions of
the proton in the
CTEQ5M set. Compared to the previous generation of distributions, such as
CTEQ4M, the most noticeable changes are in the difference of $\bar{u}$ and $%
\bar{d}$ quarks, due to the influence of the new data of E866, NMC, and CDF
W-lepton asymmetry. Fig.\ \ref{fig:dbarOubar} and Fig.\ \ref{fig:DbarMUbar} show
the combinations $\bar{d}/\bar{u}$ and $\bar{d}-\bar{u}$ respectively, in
each case comparing CTEQ5M with CTEQ4M. 
These parton distributions give excellent fits to about 1000 data points of
DIS, 140 of DY, and 57 of jet experiments. We shall bypass plots showing the
excellent fit to the well-known data sets; and focus on comparisons to data
that are new or have changed since the CTEQ4 analysis \cite{CTEQ4}. Fig.\ \ref
{fig:NmcRatio}, Fig.\ \ref{fig:Wasym}, and Fig.\ \ref{fig:DYasym} show
comparisons of NLO QCD calculations based on the CTEQ5M parton distributions
to the experimental data of NMC on the DIS deuteron to proton ratio, 
of CDF in the W-lepton asymmetry, and of E866 on the Drell-Yan
deuteron to proton ratio respectively. 
Excellent agreement is observed in all cases. There is no obvious need for a
different treatment of the deuteron data as suggested in Ref.\ \cite
{BodekYang}, although we find that the alternative scenario is also allowed
by the global analysis. 

The shifts in the quark distributions, plus the influence of the new data on
inclusive jets, also results in some shift in the gluon distribution from
CTEQ4M. Fig.\ \ref{fig:Glu4m5m} shows the CTEQ5M $G(x,Q)$ compared to CTEQ4M
at $Q=2,5,$ and $80$ GeV. These functions are uniformly scaled by $%
x^{-1.5}(1-x)^{3}$ in order to make the differences both at large and small $%
x$ visible on a linear scale. One can see that the difference between the two
distributions diminishes as $Q$ increases as the result of QCD evolution. This
well-known feature of parton distributions has many phenomenological 
consequences.  The bottom plots in Figs.\ \ref{fig:JetDzero} and
\ref{fig:JetCDF} show the measured D0 and CDF inclusive jet production
cross-sections, compared to NLO QCD calculations based on CTEQ5M (and
CTEQ5HJ, to be described later), and the top plots show the same in the
ratio form, Data/Theory. (The cross-section plot is more appropriate for
comparing the same data to different theories, as we will do later; we scale
the cross-section by $p_{t}^{7},$ so that it becomes practical to show the
comparison on a linear scale. The ratio plot is used often in experimental
papers.) \ The data are systematic error limited in most regions, except at
very large $p_{T}.$ The known correlated systematic errors, which constrain
the shape of the differential distribution, are incorporated in the global
fit. 
The normalization factors in these comparison plots are 1.04 for D0 and 1.00
for CDF. This difference in normalization factor is consistent with a known
3--5\% difference in their luminosity calibration. We will return to more 
discussions (and comparison) of these jet data in a latter section on the
CTEQ5HJ parton distribution set.

One obvious question about our determination of $G(x,Q)$ is whether we
sacrificed useful information on the gluon distribution by leaving out the
direct photon data in the analysis? \ The answer is no, due to the very
large theoretical uncertainties discussed in Sec.\ \ref{sec:GlobalIssue}.\
This situation can be illustrated by using our new parton distributions to
calculate direct photon cross-section in the existing theoretical frameworks
and comparing them with available data. Fig.\ \ref{fig:Wa70} shows the
comparison of the WA70 data with a NLO QCD calculation, using the CTEQ5M
parton distributions. A scale parameter $\mu =p_{T}/2$ is used. A
normalization factor of 1.08 is found to bring about a perfect agreement
between theory and experiment. This is not surprising since the CTEQ5
distributions are not so different from the previous generation of CTEQ4 and
MRSR distributions, which fit the WA70 data well. It is known, of course,
that the straight NLO QCD result falls well below the recent E706 data, at
somewhat higher energies. The introduction of initial state parton $k_{T}$
broadening effect can account for the difference \cite{CTEQdph}, although the
implementation of this effect is phenomenological and model dependent. Fig.%
\ref{fig:E706} shows the comparison of E706 data with a NLO QCD calculation,
with Gaussian $k_{T}$ broadening by an amount (1.2--1.3 GeV) determined from
differential distributions measured in the same experiment, cf.\ \cite
{E706,CteqE706}. The same scale parameter $\mu =p_{T}/2$ is used and no
other parameters are adjusted. The agreement is seen to be perfect. One
should note, however, if $k_{T}$ broadening is similarly introduced in the
comparison with WA70 data, using an amount seen in the WA70 di-photon
momentum imbalance say, then the agreement shown in the previous plot, Fig.%
\ref{fig:Wa70}, would no longer hold. Therefore, as discussed in Sec.\ \ref
{sec:GlobalIssue}, until the relevant experimental and theoretical issues
are resolved, direct photon data cannot be unambiguously utilized to
determine the gluon distribution.

\textbf{Large-x gluons and the CTEQ5HJ set:}

The CDF RunIA inclusive jet production data \cite{CDFIa} stimulated much
interest in physics at large $x$, in particular the possible range of the
gluon distribution in that region. The CTEQ4HJ \ parton distribution set,
proposed two years ago \cite{CTEQ4HJ}, has served as a useful example in
investigations of various large $x$ phenomena. In a subsequent systematic
study \cite{GluUnc}, we showed that the range of uncertainty of the gluon
distribution\ is quite significant beyond $x\thicksim 0.2.$ For currently
available jet production data, CDF inclusive jet $p_{T}$ distribution, 
as well as the CDF and D0 di-jet mass $m_{jj}$ distributions \cite
{DiJets}, continue to show a rise of the cross-section above the NLO QCD
calculations based on conventional parton distributions, at large $p_{T}$
and $m_{jj}$ respectively.\footnote{%
Due to the size and interpretation of current experimental errors, whether
this observed trend in each of the two experiments 
is statistically significant may be open to question.} It is
therefore desirable to update the CTEQ4HJ parton distribution set, to
complement the new CTEQ5M. This updated set is designated \textbf{CTEQ5HJ}.
It gives almost as good a global fit as CTEQ5M to the full set of data on
DIS and DY processes, with only marginally higher overall $\chi ^{2},$ and
has the feature that the gluon distribution is significantly enhanced in the
large $x$ region, resulting in improved agreement with the observed trend of 
jet data at high momentum scales mentioned
above. The existence of excellent fits of this kind again
serves to illustrate the fact that the large $x$ region remains a fertile
ground for further experimental exploration and theoretical development. Fig.%
\ref{fig:HJvsM} shows the comparison between the gluon distributions of
CTEQ5HJ and CTEQ5M at 2, 5, and 80 GeV. Due to the feature of QCD evolution
mentioned earlier, the large difference of the two distributions at low
$Q$ represents the amplified effect of fitting jet data at an energy scale 
greater than 40 GeV at the Tevatron.  The dashed lines in the bottom plots
in Figs.\ \ref{fig:JetDzero},\ref{fig:JetCDF} show the comparison of NLO QCD
calculations based on CTEQ5HJ with D0 and CDF data; and in the top plots,
they are normalized to the calculation based on CTEQ5M, along with the
relevant data. In Fig.\ \ref{fig:HJvsData}, we collect the ratio plots for the
two experiments together; both sets of data are normalized to NLO QCD
calculation based on CTEQ5HJ. This plot shows that CTEQ5HJ accounts well for
both data sets, and that the two sets are in quite good agreement with each
other. Note that experimental systematic errors are not included in this
plot; and a relative normalization factor of 4\% between the two experiments
is used (Cf.\ the previous discussions on this factor).

\textbf{Special CTEQ5 sets for Heavy Flavor physics studies:} \ 

In applying perturbative QCD to processes in which heavy quarks play an
important role, such as charm production at HERA, the standard
renormalization and factorization schemes using zero-mass heavy-quark
partons may be inadequate. See the Appendix for a fuller discussion. For
this class of applications, we obtained the \textbf{CTEQ5HQ} set using the
ACOT scheme \cite{ACOT} which gives a more accurate formulation of charm
quark physics, valid from the threshold region ($Q\thickapprox m_{c}$) to
the asymptotic region ($Q\gg m_{c}$). The ACOT scheme consists of parton
distributions defined in the (mass-independent) \msbar\ scheme, matched with
hard cross-sections calculated using on-mass-shell (i.e.\ non-zero mass)
heavy quarks when mass-effects are non-negligible. At very high momentum
scales, it reduces to the conventional zero-mass-parton \msbar\ theory. 
For energy scales not far above the heavy quark masses, it gives a more
accurate description of the underlying physics which is close to that of
the fixed-flavor-number scheme with light flavors only \cite{ACOT,Collins98}%
. In practice, for processes included in our global analysis, only the DIS
structure functions are sensitive to the difference of the conventional
(zero-mass-parton) and the ACOT (on-mass-shell parton) schemes. Thus, in
extracting CTEQ5HQ, we used ACOT scheme Wilson coefficients in calculating
DIS structure functions, along with available \msbar\ hard cross-sections
from the literature for calculating Drell-Yan, W-, Z-boson, and jet
production processes. The CTEQ5HQ set represents an updated version of
CTEQ4HQ \cite{CTEQ4HQ}, and is similar in principle to the recent MRST \cite
{MRST} distributions, which uses a different implementation of the non-zero
mass heavy parton approach.

We find the CTEQ5HQ set gives a slightly better overall fit to the full data
sets than CTEQ5M; the difference in $\chi ^{2}$ being noticeable only in the
HERA experiments, as expected. This difference is, however, not particularly
significant since both are within experimental errors. To show the effect of
the scheme choice, Fig.\ \ref{fig:MvsHQ} compares the parton distributions
from CTEQ5M and CTEQ5HQ at $Q=5$ GeV. The differences for $c(x,Q)$ and $%
G(x,Q)$ are surprisingly small; but for $u(x,Q)$ and $d(x,Q)$ they are quite
noticeable in the small $x$ region.\footnote{%
The differences for $G(x,Q)$ and $c(x,Q),$ in particular, are much smaller
than previously found between CTEQ4M and CTEQ4HQ. An important factor is the
different choices of charm quark mass $m_{c}$: $1.6$ GeV for CTEQ4, and the
more up-to-date $1.3$ GeV for CTEQ5. \cite{PDG}\ This value marks the
starting scale for radiatively generating the charm distribution. The
evolution of $c(x,Q)$ is fairly rapid in the threshold region.} This
underlines the non-trivial coupling between the various flavors when
theoretical or experimental input to the global analysis is varied.
Expectations based on direct correlations do not always hold. Given the
different treatment of charm quark mass in extracting parton distributions
in different analyses, and by various groups, an important practical
question is: \emph{how much error is incurred if these parton distributions
are applied ``incorrectly'', by convoluting them with hard cross-sections
calculated in a different scheme}?\ To answer this question quantitatively,
we evaluated the nominal $\chi ^{2}$ value of the full data sets used in our
global analysis by mixing CTEQ5M parton distributions with hard
cross-sections used in CTEQ5HQ extraction and vice versa. The $\chi ^{2}$
increased by 600 for around 1000 DIS data points (but changed little for
other processes). This big increase in $\chi ^{2}$ highlights both the
accuracy of current DIS data and the importance of maintaining consistency
in applying parton distributions in quantitative QCD calculations. Mixed use
of parton distributions defined in different schemes is clearly
unacceptable, even if the distributions may look rather similar in graphical
displays.

Although we did not use data from charm production in DIS for the CTEQ5HQ
analysis, because of the theoretical and experimental problems mentioned in
Sec.\ \ref{sec:GlobalIssue}, it is interesting for comparison purposes, 
to evaluate
an effective ``$F_{2}^{c}$'' based on the order $\alpha _{s}$ formulas in
the ACOT scheme \cite{ACOT,CTEQ4HQ} and check it against the existing
measurement. Fig.\ \ref{fig:HQvsExp} shows the result: the agreement is good. 

In some heavy quark applications, 
various authors prefer to use fixed-flavor-number
schemes, with the number of quark-partons fixed at either 3 or 4, for all
momentum scales. For these application, we present the \textbf{CTEQ5F3(4)}
sets in the fixed-3(4)-flavor scheme which treat charm (bottom) quarks as
heavy particles, not partons. The CTEQ5F3(4) sets are updated versions of
CTEQ4F3(4) \cite{CTEQ4HQ}. They are similar in spirit to the GRV parton
distributions, in particular the GRV98 set \cite{GRV98}. Whereas the
fixed-flavor-number scheme is appropriate for certain applications, such as
charm/bottom production at energy scales not far above the threshold; it is
obviously inappropriate for processes in which charm/bottom quarks play a
similar role as the light quarks, such as inclusive jet production at hadron
colliders. Thus, the range of applicability of these distributions is much
more restricted. 
%


\section{Comparison with Other Parton Distributions}

\label{sec:Compare}

To compare the CTEQ5 parton distributions with other recent parton
distribution sets, it is important to take into account not only possible
differences in input data sets and analysis procedures, but also the choices
of renormalization and factorization schemes. As already mentioned in the
description of the CTEQ5 global analysis, in addition to the familiar
differences between \msbar\ and DIS schemes, the more refined recent parton
distribution sets are also distinguished by their choice of the scheme for
treating heavy quarks.

The MRS group adopted a new procedure for treating charm quark mass effects
in DIS processes in their MRST (MRS98) analysis \cite{MRST} by applying the
method of Ref.\ \cite{RobTho97}. The conventional zero-mass formalism is
used for the other processes. This procedure is similar to that used for
CTEQ4HQ and CTEQ5HQ, although the method of \cite{RobTho97} does differ from
that of \cite{ACOT,Collins98} in the specifics of treating the mass effects (see
below). With this in mind, we first show in Fig.\ \ref{fig:HqMrsPdf} an
overview of the comparison between the CTEQ5HQ parton distributions and
those of MRST. The most striking difference is in the charm distribution;
although less obvious differences in the other flavors are also present. \
Fig.\ \ref{fig:ChmCtqMrs} makes clear that the difference in $c\left(
x,Q\right) $ spans the entire $x$ range. This difference can be attributed
to: (i) the different choice of $m_{c}$ -- $1.30\,$GeV for CTEQ5 versus $%
1.35\,$GeV for MRST; and (ii) the residual difference in the procedure of
treating charm mass effects in the Wilson coefficients. This does not
directly affect the phenomenology of charm production, since, when used in
conjunction with the appropriate Wilson coefficients, both reproduce well
the measured physical DIS structure functions, including ``$F_{2}^{c}".$
Cf.\ Fig.\ \ref{fig:HQvsExp} above and Ref.\ \cite{MRST} respectively. 

Of more phenomenological interest is the comparison of the gluon
distribution in the CTEQ and MRST analyses, because of its implications for
future high energy processes. On this issue, the difference due to the
choice of scheme is completely overshadowed by that due to the choice of
experimental input: to complement the DIS constraints in determining $%
G(x,Q), $ we used the inclusive jet data of CDF and D0, as discussed above;
whereas MRST relied on direct photon production results of WA70, applying a
range of $k_{T}$ broadening corrections using the E706 data as a constraint.
These experiments affect directly the determination of $G(x,Q)$ in the medium
to large $x$ region.
Fig.\ \ref{fig:GluCtqMrs} shows the comparison of $G(x,Q)$ from CTEQ5M and
CTEQ5HJ with those of MRST at $Q=5$ GeV. The significant difference observed
can be readily understood in terms of the inputs.

The large range of variation between the MRST sets in the region around 
$x\sim 0.25$
reflects the freedom of choice of the $k_{T}$-broadening parameter $\langle
k_{T}\rangle $ which produces a very significant correction factor to the
theoretical cross-section (recall this factor needs to be of the order of $%
2\sim 3$ for E706 to agree with data), in addition to the well-known large
scale dependence for NLO QCD predictions \cite
{AurencheOrg,CTEQdph,Aurenche98}. For a detailed discussion of the choices
made to obtain this range, see Ref.\ \cite{MRST}. The much narrower 
apparent range seen between the two CTEQ5 sets in this $x$ span is due to the
constraints on the shape of $G(x,Q)$ imposed by the inclusive jet
cross-section (which has rather stable NLO QCD theory predictions) and 
the requirement of best fit for the CTEQ5M and CTEQ5HJ conditions 
(with no attempt being made to explore the possible range as did in 
Ref.\cite{GluUnc}).
The MRST-G$\uparrow $ (MRS98-2 in the figure) set uses WA70 data with zero $%
k_{T}$ broadening. Its $G(x,Q)$ is closest to that of CTEQ5M, as can be seen
in Fig.\ \ref{fig:GluCtqMrs}.

For the $x>0.5$ region, the wide range of variation of the CTEQ5 sets
reflects the lack of experimental constraints on $G(x,Q)$ at large $x.$ \
The convergence of the MRST gluons in this region appears to be due to choosing 
the same parametrization at large $x$ for all these sets. Finally, the
differences between the two series in the range $0.01<x<0.1$ is most likely
correlated to the differences in $0.1<x<0.6$ as the result of the momentum
sum rule constraint.

The other relevant process for this discussion is inclusive jet production.
In Fig.\ \ref{fig:DzeroCtqMrs}, we show the comparison of the D0 data with NLO
QCD calculations using the two CTEQ5 and MRST series of parton distribution
sets. The calculation is performed using the Ellis-Kunzst-Soper program \cite
{EKS} with the scale parameter $\mu =E_{T}/2$ and the jet-separation
parameter $R_{sep}=1.3$ (which is the current value favored by both CDF and
D0). For this comparison, the experimental normalization is not floated, as
done in fitting the parton distributions, for the obvious reason that the
same experimental data points cannot have many different normalizations. The
MRST curves lie considerably lower than the CTEQ5 ones, because their $%
G(x,Q) $ is much lower in the relevant $x$ range, as already seen in Fig.\ \ref
{fig:GluCtqMrs}. The corresponding comparison to the CDF data is shown in 
Fig.\ \ref{fig:CdfCtqMrs}. The significance of the observed differences must 
be assessed within the context of relevant theoretical and experimental
considerations, some of which have been discussed above. 

\section{Conclusions and Comments on Uncertainties of Parton Distributions}

\label{sec:Conclude}

As both theory and experiment improve steadily, global QCD analyses
continue to show a remarkable agreement of perturbative QCD with available
data on the wide range of hard-scattering processes and allow us to extract
the non-perturbative parton distributions with increasing accuracy. There
are, however, still many areas where more detailed theoretical and
experimental work will help to clear up current uncertainties, and allow
more precise determination of the parton structure of the nucleon. We devote
this concluding section to discussions of these areas of uncertainty.

On the theory side, the most desirable advance would be a \textbf{reliable
calculation of direct photon production} 
(especially in the $p_T$ range of fixed-target experiments), 
which could elevate the
phenomenology of this process to the same level of confidence as for DIS,
DY, and jet processes, and thereby lead to a definitive determination of the
gluon distribution. Many theorists are working on the soft-gluon resummation
corrections to the NLO QCD calculation to see if this can lead to a
quantitative theory \cite{DphResum}, accounting for the factor of 3 or more
difference between the NLO theory and experiment beyond E706 energies. \
However, this explanation of the discrepancy is not yet universally accepted 
\cite{Aurenche98}.

Considerable progress has been made on the \textbf{differentiation between $%
u $ and $d$ quarks} in the last year, as the result of complementary
information provided by several different DIS and DY measurements, as
discussed in Sec.\ \ref{sec:Expts} and \ref{sec:GlobalIssue}. However, this
analysis relies heavily on: (i) the assumption of charge symmetry (i.e.$\
f_{p}^{u(d)}=f_{n}^{d(u)}$) (which has been questioned in recent literature 
\cite{ChargeSym}; and (ii) the extraction of neutron
cross-sections from actually measured deuteron cross-sections. The size of
nuclear corrections needed to extract the neutron cross-section is still a
subject of some controversy. These corrections could affect the
determination of $d/u$, especially at large $x$ \cite{BodekYang}. We found
that, in the global analysis context, all current data can be consistently
described within the PQCD formalism with or without applying a deuteron
correction; and chose to take the simple option of not applying any such
correction. A detailed study is underway to probe this issue more
thoroughly. Such studies will clearly benefit from a better theory for
nuclear corrections. Conversely, better phenomenological analyses of the
existing abundant data could provide useful input to the study of the
nuclear effects.

There has been little advance in the unambiguous determination of the 
\textbf{strange quark distribution}. The long-standing dilemma associated with
the discrepency of the strange quark distribution inferred from the di-muon neutrino data and that from the difference of neutral and charged current 
structure functions \cite{CTEQ123}
remains unresolved. This problem may be related to 
that of charge symmetry \cite{ChargeSym}.
To make real progress, the most useful
development would be measurements of physical cross-sections (or structure
functions) for charm production in neutrino-nucleon scattering, which can
then be incorporated in the global analysis. If this cannot be done for
existing measurements, one hopes it will be achievable in the analysis of
the NuTeV experiment.

The \textbf{charm quark distribution} has entered the arena of 
global QCD analysis
with the availability of charm production data in neutral current
interactions, particularly at HERA. 
This has directed attention to more precise formulations of
QCD theory including massive quarks, which have been actively pursued
over the last ten years. Unfortunately, more precise formulations
necessarily lead to additional scheme dependence of the PQCD calculations,
thereby complicate the application of the parton formalism for users of parton
distributions. We briefly described some of the pertinent issues in 
Sec.\ \ref{sec:GlobalIssue} and in the Appendix. An
interesting related question is: \emph{is there a non-perturbative component
of charm inside the nucleon? }\cite{Brodsky,HarrisSmith} This
question has not yet been addressed by any of the existing global analysis
efforts -- all assume a purely radiatively generated charm distribution
which vanishes at the threshold scale. Since the charm mass is only slightly
above the nucleon mass, there is no strong argument against the existence of
an additional non-perturbative component of charm. This issue can be studied
once more abundant precision data become available.

It is universally recognized that for a wide range of theoretical and
experimental applications, it is extremely important to know the \textbf{%
range of uncertainties of the parton distributions}. The ultimate goal would
be to have parton distribution sets with a well-defined correlation matrix
for their parameters \cite{CollinsSoper}. To see what needs to be done
toward achieving this goal, it is first necessary to recognize the major
sources of uncertainties in global QCD analysis and address them
systematically.

The most obvious uncertainties are the reported experimental errors. The
non-trivial aspect of these are the correlated systematic errors. In
principle, there are standard methods to incorporate these errors, often
represented as covariance matrices, in data-fitting. Several recent attempts
and proposals have been made to pursue this approach \cite{Gielerr}. In
practice, since only a limited number of experiments present information on
correlated errors, the input data sets for the global analysis are much more
restricted than required to determine the different parton flavors. In
addition, this task is much more complex than appears on the surface,
because: (i) it is known that the standard covariance matrix method is not
robust under certain conditions \cite{CovMtrxProb} and can lead to
pathological results\footnote{%
We have actually encountered this problem in our global fits
involving jet data, as have the CDF and D0 collaborations in their efforts
to refine the systematic errors.}, and (ii) the diversity of experiments
involved in a global analysis, and the non-uniform information they provide,
can easily vitiate some of the essential assumptions underlying the statistical
analysis method.

Theoretical uncertainties that affect the global analysis are much less
obvious and much harder to quantify than the experimental errors. The
magnitudes of the uncertainties due to higher-order effects,
scale-dependence, soft-gluon resummation, higher-twist effects, nuclear
(deuteron) corrections, etc., vary widely from process to process, and from
one kinematic region to another. Thus, while the uncertainties of NLO
calculations of DIS and DY processes are known to be under control (except
near the boundaries of the kinematic region), and those of inclusive jet
cross-section are also stable, the same is far from true for direct photon
production (at $p_T$ values of most available data) and for 
heavy quark production in hadron collisions. These
uncertainties have to be dealt with on a case by case basis, using the most
up-to-date knowledge of the specific process.

Last, but by no means least, there are hidden uncertainties associated with
the choice of functional forms for the non-perturbative initial parton
distributions. Although the parameters in these functions are determined by
comparison with experiment, the choice of functional form introduces
implicit correlations between the parton distributions at different $x$
ranges. We have encountered this hidden correlation often in our
investigation of the range of variations of the gluon distribution in
previous and current CTEQ analyses. The simpler the functional form 
(or the more economical the parametrization),%
the more rigid is the implied correlation.
\footnote{ 
A good illustration of this is the behavior of $\bar{d}/\bar{u}$ in the 
range beyond $x=0.2$ as seen in Fig.\ \ref{fig:dbarOubar}. In all
previous parton distribution sets, represented by the CTEQ4M curve in
this plot, it was determined essentially by the functional form chosen, 
with only one experimental anchor
point at x=0.18 from the NA51 measurement.  The dramatic turn around,
seen in the CTEQ5M curve, is brought about by the more extensive data set of
E866 which forced a change of the function form.}
To
reduce this undesirable correlation, one cannot, however, indiscriminately
increase the degrees of freedom of the parametrization. If there are not enough
experimental constraints to determine the parameters, one will get
unpredictable artificial behavior of the parton distributions that is not
related to the experimental input. We have also encountered examples of this
kind in the course of our analyses. Only as more precise experimental data
become available for more processes, does it become possible to refine the
parametrization in a progressive manner.

The presence of uncertainties of the second and third kind has important
implications on efforts to quantify the implications of experimental
systematic error on parton distribution analysis, because both uncertainties
are of a highly correlated nature and all three are inextricably intertwined.

In the CTEQ series of global QCD analyses \cite
{CTEQ123,CTEQ4,CTEQ4HQ,GluUnc,CtqHera}, we try to assess the current
knowledge of the parton distributions keeping all the above sources of
uncertainties in perspective, and make the best educated estimates on the
uncertainties as possible. The global analysis of parton distributions is
yet far from being an exact science, due to its complexity and comprehensive
scope. However, the steady progress that has been achieved clearly
demonstrates that vigorous pursuit of the open problems summarized above
will continue to improve our knowledge of the parton structure of hadrons,
and pave the way for advances in all fronts in elementary particle physics.

\vspace{3ex}
\textbf{Note added in proof:}
Until recently, CTEQ global analyses used QCD evolution codes
which yield slightly different results compared to those eventually
arrived at by a 1995/96 HERA working group.%
\footnote{Cf.\ J.\ Bluemlein et al., Proc.\ of 1995/96 HERA Physics
Workshop, eds.\ G.\ Ingelman et al., Vol.\ 1, p23; hep-ph/9609400.}
This was due to certain numerical approximations adopted for the NLO 
evolution.  
The approximations have since been eliminated,
so that the differences between our results 
(based on an $x$-space method) and that of the HERA group based on 
the same method are now much smaller than the differences
between their $x$-space and moment-space methods. As pointed out by the
HERA working group, the latter differences represent a
measure of the intrinsic uncertainty of these perturbative calculations.  
We have repeated the initial CTEQ5 global fits using the improved code.  
Although there are slight shifts in some parton distributions in some 
kinematic regions, all results described in this paper are independent of 
such small shifts. In particular, because the fits to data are of identical
quality, comparisons of theory and experimental data are indistinguishable.
For the same reason, the differences in current physical calculations due 
to the two versions of PDF's are small -- certainly
insignificant compared to those due to the many, much larger, 
uncertainties in the global analysis, as discussed above.  
The relative sizes of these differences will be described in forthcoming 
studies on the effect of parton distribution uncertainties on precision electro-weak and QCD phenomenology at the Tevatron and LHC.

\section*{Appendix:
Renormalization and Factorization Schemes}

\label{Scheme}

In the presence of heavy flavors, perturbative QCD becomes more involved
than commonly formulated because the heavy quark masses, $m_{i}$, appear as
extra scales in the problem. The magnitude of the \emph{hard scale} $Q$ of
the physical process relative to $m_{i}$ is an important determining factor
in what is the appropriate \emph{renormalization and factorization scheme}
to adopt. For a concise review of the relevant issues, see Ref.\ \cite
{WKTdis97}; for a rigorous theoretical treatment of the problem, see Ref.\ 
\cite{Collins98}. The following discussion applies to both charm and bottom
quarks, but for definiteness we shall focus on charm.\ 

Most universally used parton distribution functions in the past have been
generated using the conventional 4/5-quark-flavor scheme, using zero-mass hard
cross-sections for all active flavors. The same is true for practically all
the popular Monte Carlo programs used in data analyses. The implicit
assumption in this practice is that $Q\gg m_{c}$, which is obviously not a
good approximation if the application includes the region $Q\gtrsim m_{c}.$
In contrast, the fixed ``three-flavor scheme'' (in which the charm quark
never enters as a parton) has been used in many LO \cite{GRStrat} and NLO 
\cite{SBLeiden} heavy-quark production calculations, as well as in the GRV
parton distribution determination \cite{GRV94,GRV98}. The implicit
assumption here is that the charm quark always behaves like a heavy
particle, irrespective of the physical energy scale $Q$. Although this
represents the correct physical picture in the threshold region, it is
clearly inappropriate in the asymptotic region $Q\gg m_{c}.$ \ In recent
years, a unified approach (the on-mass-shell, or ACOT scheme \cite
{ACOT,Collins98,WKTdis97,KS98,ASTW}) which incorporates non-zero
charm-parton mass effects near threshold, and which contains the above two
cases as distinct limits has been formulated. Stimulated by HERA data on
charm production \cite{H1C,ZEUSC}, variants of this approach have also been
proposed and used in the recent literature \cite{MRST,RobTho97,GrecoEtal}.
Recently, Collins has provided a rigorous basis for the ACOT scheme by
establishing a generalized factorization theorem including the heavy quark
masses, to all orders of PQCD \cite{Collins98}.

The basic ideas of the on-mass-shell scheme are relatively simple: 
(i) in the region below and just
above the charm threshold ($Q\lesssim m_{c}),$ one adopts the natural
three-flavor scheme; (ii) somewhere past the charm threshold region (i.e.\
for $Q>m_{c}$), one switches to a four-flavor scheme in which the infrared
unsafe factors involving $\ln (Q/m_{c})$ are resummed into charm parton
distribution (or fragmentation) functions, while \emph{keeping} the
remaining infrared safe $m_{c}$ dependence;\footnote{%
Broadly speaking, this is the main difference between this scheme and the
commonly understood ``\msbar'' scheme, in which the entire charm mass
dependence is dropped before the collinear singularities are factored
into parton distribution/fragmentation functions.} 
and (iii) the transition from the 3-flavor to the 4-flavor scheme can
be carried out anywhere over a fairly large region in which both schemes
apply, provided appropriate matching conditions between the two schemes are
implemented. The resummation of the infrared unsafe factors enables this
approach to reproduce the conventional zero-mass parton formalism
asymptotically (i.e.\ for $Q\gg m_{c})$, while keeping the infrared safe $%
m_{c}$ effects ensures that it reproduces the fixed-three-flavor scheme
results near threshold. If the four-flavor scheme part is defined with \msbar%
\ subtraction, the parton distributions satisfy evolution equations with the
same mass-independent evolution kernel as in the conventional picture. The
hard scattering cross-sections, however, will be different from the
conventional ones in the literature if infrared safe $m_{c}$ effects are
retained in the transition region for accuracy. Furthermore, because of
inherent approximations in the perturbative approach, there are different,
but equally valid, ways to include the quark mass effects, as exemplified by
the differing results on DIS used in Refs. \cite
{ACOT,Collins98,WKTdis97,ASTW} on the one hand, and Refs.\cite{MRST,RobTho97}
on the other.

In principle, the new on-mass-shell scheme is the natural one to
use for global analysis, because of its relative generality and its special
relevance to the charm production process. However, aside from the DIS
process, no NLO hard cross-section calculation has yet been carried out in
the more general scheme. Thus, both in the previous CTEQ study, resulting in
the CTEQ4HQ distributions \cite{CTEQ4HQ}, and in the recent MRST analysis 
\cite{MRST}, non-zero-mass hard cross-sections for the DIS structure
functions are employed along with zero-mass hard cross-sections for the
other processes. This procedure appears to mix two different ways for
calculating hard cross-sections. It can however be justified in practice, if
the errors due to the zero-mass approximation are negligible over the
kinematic ranges of the relevant processes included in the analysis. This is
clearly the case for hadron collider processes (W-, Z-, and jet-production)
where the relevant momentum scale is always far greater than $m_{c}.$ For
processes involving momentum scales comparable to, or not far greater than $%
m_{c},$ the zero-mass parton approximation can still be acceptable if the
contribution from the so-called (heavy) flavor-creation (FC) subprocesses is
sub-leading, because mass effects are most significant in these subprocesses 
\cite{Collins98}. Thus it is more important to keep the mass in the DIS
calculation (where FC comes in order $\alpha _{s}$), especially for
observables involving charm particles in the final state (for which order $%
\alpha _{s}$ is in fact the leading order at a scale $Q\gtrsim m_{c}$), than
in the DY calculation (where FC only comes in at order $\alpha _{s}^{2}$).

We shall not be concerned about charm particle production in DIS in this
study, since both experimentally and theoretically, the only well-defined
physical quantities are $F_{2}^{D,D^{\ast }};$ but it is not possible at
present to use $F_{2}^{D,D^{\ast }}$ to place useful constraints on parton
distributions because they also depend on the poorly known fragmentation
functions of $D,D^{\ast }$.\footnote{%
In four-flavor schemes, including on-mass-shell ones described above, the
commonly discussed ``structure function for charm production'', $F_{2}^{c},$
is not a well-defined theoretical quantity. Naive expressions written down
for ``$F_{2}^{c}$'' will not be infrared safe -- they contain large
logarithms of the same kind which are resummed into charm parton
distributions. These are absorbed into fragmentation functions in $%
F_{2}^{D,D^{\ast }},$ making the latter well-defined theoretically.} (cf.\
Sec.\ \ref{sec:GlobalIssue}) The remaining issue is then, should one always
adopt the on-mass-shell formalism for treating the total inclusive DIS
structure functions $F_{2,3}^{total}$ in the global analysis, knowing that
leading order contributions to $F_{2,3}^{total}$ is of order $\alpha
_{s}^{0} $ and the FC contribution is of order $\alpha _{s}^{1}$ near the
threshold region where mass effects are important. From the theoretical
point of view, the answer appears to be yes -- since the formalism already
exists, and it is desirable to be as precise as possible. However, switching
to the on-shell formalism does bring about practical complications: (i) to
use the resulting parton distributions, one needs to convolute them with the
matching Wilson coefficients (at least for DIS processes) which are not
incorporated in most applications; and (ii) as we show in Sec.\ \ref{sec:CTEQ5}%
, different implementations of the (in principle equivalent) on-mass-shell
scheme actually result in quite non-negligible differences in DIS structure
function calculations at the NLO accuracy, thus users of these distributions
need to have multiple versions of hard-cross-sections to match the
corresponding parton distributions from different groups.

These considerations underlie our decision to update the conventional zero-parton-mass distributions in the form of CTEQ5M,D,L,HJ sets, which can be
used with standard hard cross-section formulas to produce accurate PQCD
calculations of all processes, along with the on-mass-shell CTEQ5HQ set which is
needed for applications involving heavy quark final states. For these latter
class of applications, we also include CTEQ5F3/4 sets in the 3/4
fixed-flavor-number schemes which are the appropriate ones to use with
existing calculations of heavy quark production processes done in these
schemes. From the discussions above, it should be clear that, in fact,
CTEQ5HQ can be regarded as the most general among all these sets, since the on-shell scheme contains the other schemes as limiting cases. 
However, in order to produce precise calculations
for DIS structure functions, CTEQ5HQ distributions must be matched with
Wilson coefficients calculated in the ACOT scheme.

\section*{Acknowledgement}

We thank many of our CTEQ colleagues,
in particular John Collins, Dave Soper, and Harry Weerts, 
for their valuable input to both experimental
and theoretical considerations relevant to the global QCD analysis project.




\input cteq5.fig		

\end{document}